# A New Algorithm for Maximum Likelihood Estimation in Gaussian Graphical Models for Marginal Independence

**Mathias Drton**
Department of Statistics
University of Washington
Seattle, WA 98195-4322

**Thomas S. Richardson**
Department of Statistics
University of Washington
Seattle, WA 98195-4322

**Abstract**

Graphical models with bi-directed edges ($\leftrightarrow$) represent marginal independence: the absence of an edge between two vertices indicates that the corresponding variables are marginally independent. In this paper, we consider maximum likelihood estimation in the case of continuous variables with a Gaussian joint distribution, sometimes termed a covariance graph model. We present a new fitting algorithm which exploits standard regression techniques and establish its convergence properties. Moreover, we contrast our procedure to existing estimation algorithms.

## 1 INTRODUCTION

Graphical models are commonly based on undirected graphs, DAGs, or chain graphs in which the absence of an edge between two vertices indicates some conditional independence between the associated variables. However, there has also been interest in graphical models in which the variables associated with two non-adjacent vertices are marginally independent. These models for marginal independence may naturally be represented with bi-directed edges ($\leftrightarrow$) via a natural extension of d-separation. These models correspond to the special case of Richardson's and Spirtes' (2002) *ancestral graph* models where the ancestral graph has bi-directed edges only. The case of jointly Gaussian variables has also been termed *covariance graph* models by Cox and Wermuth (1993, 1996) who use dashed lines rather than bi-directed edges. Besides being interesting models in their own right, graphical models for marginal independence are also interesting in the context of DAGs since certain DAGs with hidden variables induce marginal independences amongst the observed variables that can be represented by a bi-directed graph (see Section 2).

For undirected graphs, DAGs and chain graphs, parameter learning procedures are well developed, see e.g. Lauritzen (1996) or Whittaker (1990), and many methods are implemented in Edwards' (2000) software package MIM. This is not the case, however, for graphical models for marginal independence and covariance graph models in particular. For instance, MIM does not permit maximum likelihood (ML) estimation in covariance graph models but permits fitting only by Kauermann's (1996) heuristic method based on a "dual likelihood", compare Edwards (2000, §7.4).

This article presents a new iterative algorithm for ML estimation in covariance graph models. In Section 2 we describe and motivate graphical models for marginal independence in general, and in Section 3 we turn to the Gaussian case of covariance graph models. In particular, we review and critique Anderson's ML estimation algorithm. In the core Section 4 we present our new algorithm which converges to a solution of the likelihood equations for almost every value of the observations. In Section 5 we show the estimates for an example data set and in Section 6 we outline future extensions to our algorithm.

## 2 GRAPHICAL MODELS FOR MARGINAL INDEPENDENCE

Suppose that we observe the set of variables $V$ in the random vector $X = (X_i \mid i \in V)$. Let $G = (V, E)$ be a graph with the variable set $V$ as vertex set and the edge set $E \subsetneq V \times V$ consisting exclusively of bi-directed edges $(i, j)$, $i \neq j \in V$, denoted by $i \leftrightarrow j$.

### 2.1 GLOBAL MARKOV PROPERTY FOR BI-DIRECTED GRAPHS

In the bi-directed graph $G$, a path between vertices $i$, $j \in V$ is said to *m-connect* given $S \subseteq V$ if there is a path between $i$ and $j$ on which every non-endpoint vertex is in $S$. Disjoint non-empty sets of vertices $A$ and $B$ are *m-connected* given $S$ in $G$ if for some $i \in A$ and $j \in B$ there exists an m-connecting path between $i$ and $j$ given $S$ in $G$. Otherwise, $A$ and $B$ are *m-separated* given $S$ where $S$ is allowed to be empty. The distribution of $X$ satisfies the *global Markov property* for $G$ if $X_A \perp\!\!\!\perp X_B \mid X_S$ holds whenever $A$ is m-separated from $B$ given $S$. Here, $X_A = (X_i \mid i \in A)$, etc. Note that the global Markov property implies the *pairwise Markov property* consisting of the marginal independences $X_i \perp\!\!\!\perp X_j$ for all $i \not\leftrightarrow j$. If $X$ has a Gaussian distribution then the pairwise Markov property also implies the global Markov property but this does not need to hold for an arbitrary probability distribution.

In the graph shown in Figure 1(a), the path $1 \leftrightarrow 3 \leftrightarrow 4 \leftrightarrow 2$ m-connects 1 and 2 given $Z = \{3,4\}$, but 1 and 2 are m-separated given any proper subset of $\{3,4\}$. The global Markov property for this graph requires that $1 \perp\!\!\!\perp \{2,4\}$ and $2 \perp\!\!\!\perp \{1,3\}$.

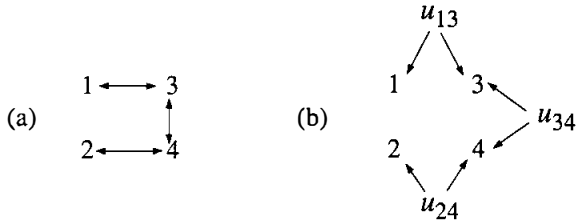

Figure 1: (a) Bi-directed Graph. (b) DAG with Hidden Variables $u_{13}$, $u_{34}$, $u_{24}$ (Section 2.2).

In a bi-directed graph, there are no directed paths, and every non-endpoint vertex on a path is a *collider*, i.e. two arrow-heads point to the vertex. Therefore, the definition of m-connection given above is the natural extension of Pearl's (1988) d-connection criterion to graphs containing bi-directed edges. Richardson and Spirtes (2002) define m-connection for a larger class of graphs called ancestral graphs, which may contain directed, undirected and bi-directed edges. The definition for ancestral graphs reduces to the definition given here in the case where all edges are bi-directed.

### 2.2 RELATION TO DAG MODELS WITH HIDDEN VARIABLES

Graphical models for marginal independence can be motivated by the following consideration (see also Pearl and Wermuth, 1994). Suppose that there is DAG $D$ with vertex set $V \cup U$, where the variables in $V$ are observed, and those in $U$ are unobserved. Suppose further that observed variables $i \in V$ have no children in the graph, i.e. $\text{ch}_D(i) = \{j \in V \cup U \mid i \to j\} = \emptyset$. Models of this kind are used in psychology and the social sciences, see e.g. Bollen (1989, §6).

From the DAG $D$, construct the induced bi-directed graph $G(D)$ over $V$ by including the bi-directed edge $i \leftrightarrow j$, $i,j \in V$, if $\text{an}_D(i) \cap \text{an}_D(j) \neq \emptyset$, where $\text{an}_D(i) = \{j \mid j \to \ldots \to i \text{ or } j = i\}$ are the ancestors of $i$ in $D$. Note that $\text{an}_D(i) \cap \text{an}_D(j) \subseteq U$. It then follows as a special case of Theorem 4.18 in Richardson and Spirtes (2002) that $G(D)$ represents the Markov structure induced by $D$ on the observed variables.

**Proposition 1** *Let $D$ be a DAG with vertex set $V \cup U$ such that $\text{ch}_D(i) = \emptyset$ for all $i \in V$, and let $G(D)$ be the bi-directed graph with vertex set $V$ defined above. Then for any disjoint sets $A, B, S \subseteq V$, with $A, B \neq \emptyset$,*

$A$ *and* $B$ *are d-separated given* $S$ *in* $D$
$\Leftrightarrow A$ *and* $B$ *are m-separated given* $S$ *in* $G(D)$.

Figure 1(b) shows a DAG $D$ with $G(D)$ equal to the graph shown in Figure 1(a).

DAGs with hidden variables that satisfy the conditions of Proposition 1 induce an independence structure over the observed variables that can be represented by a bi-directed graph. However, further (non-independence) restrictions can hold in the hidden variable model such that it is only a submodel of the bi-directed graph model (see Richardson and Spirtes 2002, §7.3.1, §8.6).

### 2.3 MARKOV EQUIVALENCE RESULTS

A bi-directed graph is Markov equivalent to some DAG with the same set of vertices if the respective global Markov properties yield the same conditional independences. Pearl and Wermuth (1994) state the result given in Proposition 2. In Proposition 3, we give the natural converse.

**Proposition 2** *Let $G$ be a bi-directed graph with vertex set $V$. Then (i) there is a DAG $D$ with the same vertex set $V$, which is Markov equivalent to $G$, iff (ii) $G$ does not contain either of the following as induced subgraphs:*

$$a \leftrightarrow b \leftrightarrow c \leftrightarrow d \qquad a \leftrightarrow b \leftrightarrow c \leftrightarrow d \leftrightarrow a.$$

**Proposition 3** *Let $D$ be a DAG with vertex set $V$. Then (i) there is a bi-directed graph $G$ with vertex set $V$, which is Markov equivalent to $D$, iff (ii) $D$ does not contain an unshielded non-collider, i.e. if $a$ and $b$ are adjacent, $b$ and $c$ are adjacent, but $a$ and $c$ are not adjacent then $a \to b \leftarrow c$ in $D$.*

## 3　COVARIANCE GRAPH MODELS

Suppose now that the variables $V$ are continuous with a centered Gaussian $\equiv$ normal joint distribution, i.e. $X \sim \mathcal{N}_V(0, \Sigma) \in \mathbb{R}^V$ where $\Sigma = (\sigma_{ij}) \in \mathbb{R}^{V \times V}$ is the unknown positive definite covariance matrix. The normal distribution of $X$ is pairwise and globally Markov for the bi-directed graph $G = (V, E)$ iff

$$X_i \perp\!\!\!\perp X_j \quad \Longleftrightarrow \quad \sigma_{ij} = 0 \quad \Longleftrightarrow \quad i \not\leftrightarrow j. \quad (1)$$

Let $\mathbf{P}(V)$ be the cone of all positive definite $V \times V$ matrices and let $\mathbf{P}(G)$ be the cone of all matrices $\Sigma \in \mathbf{P}(V)$ which fulfill the linear restrictions in (1). Then the *covariance graph model* based on $G$ is the family of all normal distributions

$$\mathbf{N}(G) = \{\mathcal{N}_V(0, \Sigma) \mid \Sigma \in \mathbf{P}(G)\}. \quad (2)$$

This article considers the estimation of the unknown parameter $\Sigma$ based on a sample of i.i.d. observations $X^{(k)}$, $k \in N = \{1, \ldots, n\}$, from the covariance graph model (2). The set $N$ can be interpreted as indexing the subjects on which we observe the variables in $V$. We group the vectors in the sample as columns in the $V \times N$ random matrix $Y$ which is distributed as

$$Y \in \mathbb{R}^{V \times N} \sim \mathcal{N}_{V \times N}(0, \Sigma \otimes I_N). \quad (3)$$

Here, $I_N$ is the $N \times N$ identity matrix, $\Sigma \in \mathbf{P}(G)$ is the unknown positive definite covariance matrix, and $\otimes$ is the Kronecker product. Thus the $i$-th row $Y_i = Y_i. \in \mathbb{R}^N$ of the matrix $Y$ contains the i.i.d. observations for variable $i \in V$ on all the subjects in $N$ and the $k$-th column $Y_{\cdot k} = X^{(k)}$ holds all the observations made on the subject $k \in N$. Finally, the sample size is $n = |N|$ and the number of variables is $p = |V|$.

Since our model assumes a zero mean, the empirical covariance matrix is defined to be

$$S = \frac{1}{n} \sum_{k \in N} Y_{\cdot k} Y_{\cdot k}' \in \mathbb{R}^{V \times V}. \quad (4)$$

We shall assume that $n \geq p$ such that $S$ is positive definite with probability one.

Note that the case where the model also includes an unknown mean vector $\mu$ can be treated by estimating $\mu$ by the empirical mean vector $\bar{Y} \in \mathbb{R}^V$, i.e. the vector of the row means of $Y$. The empirical covariance matrix would then be the matrix

$$\tilde{S} = \frac{1}{n} \sum_{k \in N} (Y_{\cdot k} - \bar{Y})(Y_{\cdot k} - \bar{Y})' \in \mathbb{R}^{V \times V}, \quad (5)$$

and learning of $\Sigma$ would use $\tilde{S}$ instead of $S$. However, we would have to assume that $n \geq p+1$ to ensure that $\tilde{S}$ is positive definite with probability one.

### 3.1 MAXIMUM LIKELIHOOD ESTIMATION

The density function of $Y$ with respect to the Lebesgue measure is the function $f_\Sigma : \mathbb{R}^{V \times N} \to \mathbb{R}$ given by

$$f_\Sigma(y) = (2\pi)^{-np/2} |\Sigma|^{-n/2} \times \\ \exp\Big\{ -\frac{1}{2} \sum_{k \in N} y_{\cdot k}' \Sigma^{-1} y_{\cdot k} \Big\}. \quad (6)$$

Considered as a function of the unknown parameters for fixed data $y$ this gives the likelihood function of the covariance graph model $\mathbf{N}(G)$ as the mapping $L : \mathbf{P}(G) \to \mathbb{R}$ where $L(\Sigma) = f_\Sigma(y)$. In ML estimation, the parameter $\Sigma$ is estimated by the element of $\mathbf{P}(G)$ which maximizes the likelihood $L$ or more conveniently the log-likelihood $\ell(\Sigma) = \log L(\Sigma)$.

The log-likelihood $\ell(\Sigma)$ can be expressed as

$$\ell(\Sigma) = -\frac{np}{2}\log(2\pi) - \frac{n}{2}\log|\Sigma| - \frac{n}{2}\mathrm{tr}\{\Sigma^{-1}S\}, \quad (7)$$

see e.g. Edwards (2000, §3.1). The condition $n \geq p$ implies the existence of the global maximum of $\ell(\Sigma)$ over $\mathbf{P}(G)$. This condition is not necessary in general but we are not aware of any results in the literature which provide a necessary and sufficient condition.

Besides existence, there is also the issue of uniqueness of the ML estimates, i.e. whether the likelihood has a unique local maximum which is then global. The model $\mathbf{N}(G)$ is a curved but not regular exponential family, thus, the log-likelihood need not be concave. In fact, the log-likelihood can have multiple local maxima (cf. Drton and Richardson 2002). For a large enough sample size, a multimodal likelihood seems not to arise in practice assuming the model assumptions hold but might still arise if the model assumptions do not hold (see also Cox and Wermuth 1996, p. 102f).

### 3.2 THE LIKELIHOOD EQUATIONS

The likelihood equations are the estimating equations obtained by setting the derivatives of the log-likelihood $\ell(\Sigma)$ with respect to $\sigma_{ij}$, $i = j$ or $i \leftrightarrow j$, to zero. From Anderson and Olkin (1985, §2.1.1) it follows that the likelihood equations are

$$(\Sigma^{-1})_{ij} = (\Sigma^{-1} S \Sigma^{-1})_{ij} \quad (8)$$

for $i = j$ and $i \leftrightarrow j$. The full matrix $\Sigma$ is then determined by $\sigma_{ij} = 0$ for $i \not\leftrightarrow j$.

### 3.3 ANDERSON'S ALGORITHM

Anderson (1969, 1970) studied general linear hypotheses on the covariance matrix $\Sigma$ which contain covariance graph models as a special case. In Anderson (1973), he developed an iterative algorithm to solve specifically the likelihood equations under linear hypotheses on $\Sigma$. We refer to this estimation procedure as *Anderson's algorithm*.

The iterations in Anderson's algorithm consist of solving a system of linear equations built from the current estimate of $\Sigma$. In the case of a covariance graph model, the linear equations are solved for the vector of unrestricted elements in $\Sigma$, i.e. for $\sigma = (\sigma_{ij} \mid (i,j) \in F)$ where $F = \{ij \equiv (i,j) \mid i \leftrightarrow j \lor i = j\}$, and the algorithm can be specified as follows. We denote $\sigma^{ij} = (\Sigma^{-1})_{ij}$ and define $A = A_\Sigma$ to be the $F \times F$ matrix with

$$A_{(ij,kk)} = \sigma^{ik}\sigma^{jk} \qquad \forall ij \in F,\ k \in V, \quad (9)$$
$$A_{(ij,k\ell)} = \sigma^{ik}\sigma^{j\ell} + \sigma^{jk}\sigma^{i\ell} \qquad \forall ij \in F,\ k \leftrightarrow \ell. \quad (10)$$

Furthermore, we set the $F \times 1$ vector $b = b_\Sigma$ to

$$b_{ij} = (\Sigma^{-1} S \Sigma^{-1})_{ij} \quad \forall ij \in F. \quad (11)$$

It can be shown that $\Sigma \in \mathbf{P}(G)$ solves $A_\Sigma \sigma = b_\Sigma$ iff $\Sigma$ solves the likelihood equations (8).

Anderson's algorithm updates the current estimate $\Sigma^{(r)}$ to $\Sigma^{(r+1)}$ determined by the linear equations

$$A_{\Sigma^{(r)}}\ \sigma^{(r+1)} = b_{\Sigma^{(r)}}. \quad (12)$$

Thus, a fixed point of Anderson's algorithm is a solution to the likelihood equations (8). As starting value, Anderson suggests the identity matrix, i.e. $\Sigma^{(0)} = I_V$. In the first step, his algorithm constructs the empirical estimate $\Sigma^{(1)}$ with $\sigma^{(1)}_{ij} = S_{ij},\ ij \in F$. However, neither $\Sigma^{(1)}$ nor any subsequent estimate of $\Sigma$ has to be positive (semi-) definite and thus may not be a valid covariance matrix. Moreover, at any given stage, the likelihood may decrease, and convergence of Anderson's algorithm cannot be guaranteed.

Therefore, we propose an alternative algorithm for ML estimation which constructs a sequence of estimates in $\mathbf{P}(G)$ with never decreasing likelihood which converges to a solution of the likelihood equations for almost every data matrix $y$. Note that our new algorithm only fits covariance graph models and cannot treat the wide range of models covered by Anderson's algorithm.

## 4 NEW ALGORITHM

### 4.1 THE IDEA

Let $i$ be a variable index in $V$ and set $-i = V \setminus \{i\}$. For $A, B \subseteq V$, $\Sigma_{A,B}$ denotes the $A \times B$ submatrix of $\Sigma$ and $Y_A$ denotes the $A \times N$ submatrix of $Y$. The conditional distribution of $Y_i$ given $Y_{-i}$ is

$$(Y_i \mid Y_{-i}) \sim \mathcal{N}_{\{i\} \times N}(B_i Y_{-i}, \lambda_i I_N) \in \mathbb{R}^{\{i\} \times N}, \quad (13)$$

where

$$B_i = \Sigma_{i,-i} \Sigma_{-i,-i}^{-1} \in \mathbb{R}^{\{i\} \times -i} \quad (14)$$

is the $\{i\} \times -i$ matrix of regression coefficients and

$$\lambda_i = \sigma_{ii} - \Sigma_{i,-i} \Sigma_{-i,-i}^{-1} \Sigma_{-i,i} \in \mathbb{R} \quad (15)$$

is the conditional variance. The joint density function for $Y$ can be factored into the product of the marginal density function for $Y_{-i}$ and the conditional density function of $Y_i$ given $Y_{-i}$:

$$f_\Sigma(y) = f_{(B_i,\lambda_i)}(y_i \mid y_{-i}) f_{\Sigma_{-i,-i}}(y_{-i}). \quad (16)$$

Our idea for an iterative ML estimation algorithm is then to treat $\Sigma_{-i,-i}$ as if it were known from a current feasible estimate of $\Sigma$, to estimate $B_i$ and $\lambda_i$ from the regression (13), and to then update $\Sigma$ according to (14) and (15) using the three pieces $\Sigma_{-i,-i}$, $B_i$, and $\lambda_i$. Of course, this step needs to be carried out in turn for all $i \in V$.

The subtlety with this idea is that we need to respect the restriction $\Sigma \in \mathbf{P}(G)$ when estimating $B_i$ and $\lambda_i$. If the graph $G$ was the complete graph $\bar{G}$ in which an edge joins any pair of vertices then the mapping

$$\begin{aligned}\mathbf{P}(\bar{G}) = \mathbf{P}(V) &\to (0,\infty) \times \mathbb{R}^{\{i\} \times -i} \times \mathbf{P}(-i),\\ \Sigma &\mapsto (\lambda_i, B_i, \Sigma_{-i,-i})\end{aligned} \quad (17)$$

would be bijective and the regression in (13) a standard normal regression. For a general graph $G$, this will not be the case. Fortunately, our covariance restrictions are linear and so simple that we will be able to find equivalent restrictions on the regression coefficients $B_i$ which will lead to an equivalent standard normal regression based, however, on altered variables which we name pseudo-variables.

### 4.2 AN EXAMPLE

Before we develop our idea to a generally applicable algorithm, we illustrate it by the example of the graph shown in Figure 1(a). This graph imposes the marginal independences $1 \perp\!\!\!\perp \{2,4\}$ and $2 \perp\!\!\!\perp \{1,3\}$ which imply the conditional independences $3 \perp\!\!\!\perp 2 \mid 1$ and $4 \perp\!\!\!\perp 1 \mid 2$. Thus the joint density can be decomposed as

$$f(y) = f(y_3, y_4 \mid y_1, y_2) f(y_2) f(y_1). \quad (18)$$

In this factorization the term $f(y_3, y_4 \mid y_1, y_2)$ corresponds to the bivariate seemingly unrelated regression model considered in Drton and Richardson (2002).

Here, however, we do not want to make use of the factorization (18) which holds in this special example but we wish to demonstrate how our algorithm proceeds in the general setting. Because of the symmetry of the graph we only describe the regressions $(Y_1 \mid Y_2, Y_3, Y_4)$ and $(Y_3 \mid Y_1, Y_2, Y_4)$. The remaining two regressions for $Y_2$ and $Y_4$ are the obvious analogs obtained by exchanging the indices 1 and 2, as well as 3 and 4.

### 4.2.1 Update Step For Variable 1

Let the *spouses* of $i \in V$ be $\text{sp}(i) = \{j \mid i \leftrightarrow j\}$ and the *non-spouses* $\text{nsp}(i) = V \setminus (\text{sp}(i) \cup \{i\})$. For $i = 1$, we find $\text{sp}(1) = 3$ and $\text{nsp}(1) = 24$ where the shorthand $ij$ denotes the set $\{i,j\}$. The bijection (17) suggests that we can improve our current estimate of $\Sigma$ by holding the block $\hat{\Sigma}_{234,234}$ fixed and using the regression (13) to find improved estimates of $B_i$, $\lambda_i$ and hence of $\Sigma_{1,1234} = \Sigma'_{1234,1}$. However, we need to respect the two restrictions $\sigma_{12} = \sigma_{14} = 0$. Since these are restrictions of marginal independence they do not translate into restricting some of the regression coefficients in $B_1$ to zero (as would be the case with an undirected graphical model). Instead some coefficients in $B_1$ will be a linear combination of the remaining entries in $B_1$.

More specifically, let $\beta_{ij}$ and $\beta_{ij,K}$ denote the regression coefficient for $Y_j$ in the regressions $(Y_i \mid Y_j)$ and $(Y_i \mid Y_{\{j\} \cup K})$, respectively. It follows from (14) that $B_1 \Sigma_{234,234} = \Sigma_{1,234} = (0, \sigma_{13}, 0)$. Since $B_1 = (\beta_{12.34}, \beta_{13.24}, \beta_{14.23})$ this implies

$$\beta_{13.24} \Sigma_{3,24} + (\beta_{12.34}, \beta_{14.23}) \Sigma_{24,24} = (0,0). \quad (19)$$

Thus,

$$\begin{aligned}(\beta_{12.34}, \beta_{14.23}) &= -\beta_{13.24} \Sigma_{3,24} \Sigma_{24,24}^{-1} \\ &= -\beta_{13.24}(\beta_{32.4}, \beta_{34.2}).\end{aligned} \quad (20)$$

From (20), it follows that $B_1 Y_{234} = \beta_{13.24} Z_1$ where the pseudo-variable $Z_1$ equals

$$Z_1 = Y_3 - \beta_{32.4} Y_2 - \beta_{34.2} Y_4 \in \mathbb{R}^{\{1\} \times N}. \quad (21)$$

The regression (13) then becomes

$$(Y_1 \mid Y_2, Y_3, Y_4) \sim \mathcal{N}(\beta_{13.24} Z_1, \lambda_1 I_N). \quad (22)$$

Since we hold $\hat{\Sigma}_{234,234}$ fixed it can be used to compute current estimates of the regression coefficients $\hat{\beta}_{32.4}$ and $\hat{\beta}_{34.2}$ which are plugged into (21) to yield the estimate $\hat{Z}_1$ of $Z_1$. We use $\hat{Z}_1$ in the regression (22) and find from the usual least squares formulas:

$$\begin{aligned}\hat{\beta}_{13.24} &= Y_1 \hat{Z}'_1 (\hat{Z}_1 \hat{Z}'_1)^{-1}, \\ \hat{\lambda}_1 &= \frac{1}{n} Y_1 (I_N - \hat{Z}'_1 (\hat{Z}_1 \hat{Z}'_1)^{-1} \hat{Z}_1) Y'_1.\end{aligned} \quad (23)$$

Using (20) we can compute $\hat{\beta}_{12.34}$ and $\hat{\beta}_{14.23}$ to complete the estimate $\hat{B}_1$.

With the new estimates, we update $\hat{\Sigma}$ as follows. The block $\hat{\Sigma}_{234,234}$ remains unchanged. From (14) and (15), we obtain $\hat{\Sigma}_{1,234} = \hat{B}_1 \hat{\Sigma}_{234,234}$ and $\hat{\Sigma}_{234,1} = \hat{\Sigma}'_{1,234}$. Finally, we update $\hat{\sigma}_{11}$ to $\hat{\sigma}_{11} = \hat{\lambda}_1 + \hat{B}_1 \hat{\Sigma}_{234,1}$.

### 4.2.2 Update Step For Variable 3

For $i = 3$, $\text{sp}(3) = 14$ and $\text{nsp}(3) = 2$. In regression (13), we must now respect $\sigma_{32} = 0$ which, by a similar calculation as in Section 4.2.1, translates into

$$\beta_{32.14} = -\beta_{31.24} \beta_{12} - \beta_{34.12} \beta_{42}. \quad (24)$$

Therefore the regression (13) is now

$$(Y_3 \mid Y_1, Y_2, Y_4) \sim \mathcal{N}((\beta_{31.24}, \beta_{34.12}) Z_3, \lambda_3 I_N) \quad (25)$$

with the pseudo-variables

$$Z_3 = \begin{pmatrix} Y_1 - \beta_{12} Y_2 \\ Y_4 - \beta_{42} Y_2 \end{pmatrix} \in \mathbb{R}^{\{1,4\} \times N}. \quad (26)$$

We fix $\hat{\Sigma}_{124,124}$ and compute from it the regression coefficients $\hat{\beta}_{12}$ and $\hat{\beta}_{42}$ yielding $\hat{Z}_3$ by (26). Using $\hat{Z}_3$ in the regression (25), we obtain the least squares estimates

$$\begin{aligned}(\hat{\beta}_{31.24}, \hat{\beta}_{34.12}) &= Y_3 \hat{Z}'_3 (\hat{Z}_3 \hat{Z}'_3)^{-1}, \\ \hat{\lambda}_3 &= \frac{1}{n} Y_3 (I_N - \hat{Z}'_3 (\hat{Z}_3 \hat{Z}'_3)^{-1} \hat{Z}_3) Y'_3.\end{aligned} \quad (27)$$

Using (24) we can compute $\hat{\beta}_{32.14}$ to complete the estimate $\hat{B}_3$. In the resulting update of $\hat{\Sigma}$, the submatrix $\hat{\Sigma}_{124,124}$ remains unchanged but we set $\hat{\Sigma}_{3,124} = \hat{B}_3 \hat{\Sigma}_{124,124}$ and $\hat{\Sigma}_{124,3} = \hat{\Sigma}'_{3,124}$. The remaining variance $\hat{\sigma}_{33}$ is updated to $\hat{\sigma}_{33} = \hat{\lambda}_3 + \hat{B}_3 \hat{\Sigma}_{124,3}$.

### 4.2.3 The Iteration

Figure 2 illustrates one full iteration in our algorithm in this example. The algorithm cycles in arbitrary order through the four regressions $(Y_i \mid Y_{-i})$, $i = 1, 2, 3, 4$. In Figure 2, a filled circle represents variables in the conditioning set $-i$, and an unfilled circle stands for the variable $i$ forming the response variable in the considered regression. The thick directed edges coincide with bi-directed edges in the original graph shown in Figure 1. Thin edges do not have a corresponding bi-directed edge in the original graph. Regression coefficients label the edges. It can be seen that the regression coefficients at thin edges are linear combinations of the regression coefficients at thick edges where the weights in the linear combinations are decorated with "hats" as $\hat{\beta}$ to remind that they are computed from $\hat{\Sigma}_{-i,-i}$, the block remaining unchanged in the $i$-step of the iteration. Regression coefficients without "hat" are estimated by regression on appropriate pseudo-variables.

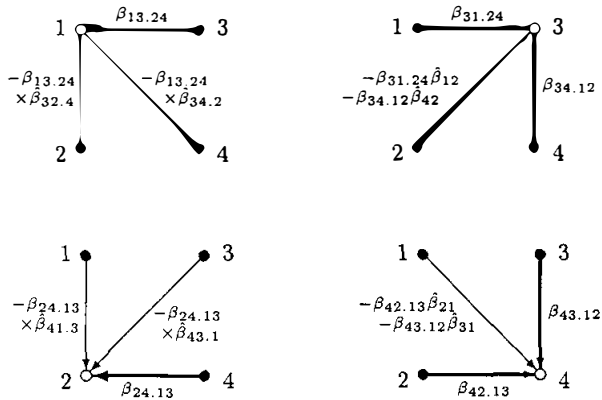

Figure 2: Illustration of New Algorithm.

#### 4.2.4 A Criticism

Our algorithm does not make use of any available likelihood factorization. For example, (18) implies that the components $\hat{\sigma}_{11}$ and $\hat{\sigma}_{22}$ of a solution to the likelihood equations must coincide with the corresponding empirical quantities $S_{11}$ and $S_{22}$, respectively. Thus the pseudo-variable regressions $(Y_1 \mid Y_2, Y_3, Y_4)$ and $(Y_2 \mid Y_1, Y_2, Y_4)$ need not be carried out, and the algorithm's convergence is sped up considerably. Hence, the algorithm may be improved by systematically employing information on which submatrices of a solution to the likelihood equations must coincide with their empirical counterparts, but this requires further work.

### 4.3 PSEUDO-VARIABLE REGRESSION

We now describe the general algorithm. Let $\hat{\Sigma}^* \in \mathbf{P}(G)$ be a feasible estimate of $\Sigma$. Suppose we wish to update $\hat{\Sigma}^*$ to a new estimate $\hat{\Sigma} \in \mathbf{P}(G)$ by setting $\hat{\Sigma}_{-i,-i} = \hat{\Sigma}^*_{-i,-i}$ and using the regression $(Y_i \mid Y_{-i})$ to obtain $\hat{\Sigma}_{i,V}$. For $A \subseteq V$, define $\mathbf{P}_A(G)$ to be the set of all $A \times A$ submatrices of matrices in $\mathbf{P}(G)$. Then a matrix $\Sigma \in \mathbf{P}(G)$ iff $\Sigma$ fulfills the two conditions

(1) $\Sigma_{-i,-i} \in \mathbf{P}_{-i}(G)$,

(2) $(B_i \Sigma_{-i,-i})_{ij} = \sigma_{ij} = 0 \quad \forall i \not\leftrightarrow j$.

Since $\hat{\Sigma}_{-i,-i} = \hat{\Sigma}^*_{-i,-i}$ condition (1) is fulfilled by $\hat{\Sigma}$ because $\hat{\Sigma}^* \in \mathbf{P}(G)$ by assumption. The remaining condition (2) can be rewritten as $B_i \Sigma_{-i,\mathrm{nsp}(i)} = 0$ and further as

$$B_{i,\mathrm{sp}(i)} \Sigma_{\mathrm{sp}(i),\mathrm{nsp}(i)} + B_{i,\mathrm{nsp}(i)} \Sigma_{\mathrm{nsp}(i),\mathrm{nsp}(i)} = 0. \quad (28)$$

Hence, the regression coefficients for the non-spouses of $i$ are linear combinations of the regression coeffi-cients for the spouses:

$$\begin{aligned} B_{i,\mathrm{nsp}(i)} &= -B_{i,\mathrm{sp}(i)} \Sigma_{\mathrm{sp}(i),\mathrm{nsp}(i)} \Sigma^{-1}_{\mathrm{nsp}(i),\mathrm{nsp}(i)} \\ &= -B_{i,\mathrm{sp}(i)} B_{\mathrm{sp}(i).\mathrm{nsp}(i)}, \end{aligned} \quad (29)$$

where

$$B_{\mathrm{sp}(i).\mathrm{nsp}(i)} = \Sigma_{\mathrm{sp}(i),\mathrm{nsp}(i)} \Sigma^{-1}_{\mathrm{nsp}(i),\mathrm{nsp}(i)} \quad (30)$$

are the regression coefficients in $(Y_{\mathrm{sp}(i)} \mid Y_{\mathrm{nsp}(i)})$. From (29), we obtain that the mapping

$$\begin{aligned} \mathbf{P}(G) &\to (0,\infty) \times \mathbb{R}^{\{i\} \times \mathrm{sp}(i)} \times \mathbf{P}_{-i}(G), \\ \Sigma &\mapsto (\lambda_i, B_{i,\mathrm{sp}(i)}, \Sigma_{-i,-i}) \end{aligned} \quad (31)$$

is bijective. Hence, for restricted $\Sigma \in \mathbf{P}(G)$ the submatrix $\Sigma_{-i,-i}$ does not restrict the range of variation of $\lambda_i$ and $B_{i,\mathrm{sp}(i)}$ in the maximization of the likelihood factor $f_{(B_i,\lambda_i)}(y_i \mid y_{-i})$ in (16). Moreover, for $\Sigma \in \mathbf{P}(G)$, the regression (13) can be rewritten as

$$(Y_i \mid Y_{-i}) \sim \mathcal{N}_{\{i\} \times N}(B_{i,\mathrm{sp}(i)} Z_i, \lambda_i I_N) \in \mathbb{R}^{\{i\} \times N}, \quad (32)$$

where the *pseudo-variables* $Z_i$ are the residuals in the regression $(Y_{\mathrm{sp}(i)} \mid Y_{\mathrm{nsp}(i)})$, i.e.

$$Z_i = Y_{\mathrm{sp}(i)} - B_{\mathrm{sp}(i).\mathrm{nsp}(i)} Y_{\mathrm{nsp}(i)} \in \mathbb{R}^{\mathrm{sp}(i) \times N}. \quad (33)$$

Since $\mathrm{sp}(i) \subseteq -i$ and $\mathrm{nsp}(i) \subseteq -i$ the regression coefficients $\hat{B}_{\mathrm{sp}(i).\mathrm{nsp}(i)}$ can be calculated from $\hat{\Sigma}^*_{-i,-i}$. These can be plugged into (33) to find estimates $\hat{Z}_i$ of the pseudo-variables. The pseudo-variable regression (32) yields new estimates $\hat{B}_{i,\mathrm{sp}(i)}$ and $\hat{\lambda}_i$ from which we can calculate $\hat{B}_{i,\mathrm{nsp}(i)}$ using (29). Thus, we can form the estimate $\hat{B}_i$ from which we can reconstruct $\hat{\Sigma}_{i,-i}$ using (14) and $\hat{\sigma}_{ii}$ using (15).

### 4.4 THE ALGORITHM

These considerations lead to the following algorithm where $\hat{\Sigma}^{(r)}$ denotes the estimated value of $\Sigma$ after the $r$-th iteration, and $\hat{\Sigma}^{(r,i)}$ is the estimated value of $\Sigma$ after the $i$-th step of the $r$-th iteration, i.e. after regressing $Y_i$ on $Y_{-i}$.

(1) Set the iteration counter $r = 0$, and choose a starting value $\hat{\Sigma}^{(0)} \in \mathbf{P}(G)$, e.g. the identity matrix $\hat{\Sigma}^{(0)} = I_V$.

(2) Order the variables in $V$ as $V = \{1, \ldots, p\}$, set $\hat{\Sigma}^{(r,0)} = \hat{\Sigma}^{(r)}$, and repeat the following steps for all $i = 1, \ldots, p$:

  (a) Let $\hat{\Sigma}^{(r,i)}_{-i,-i} = \hat{\Sigma}^{(r,i-1)}_{-i,-i}$ and calculate from this submatrix the regression coefficients $\hat{B}_{\mathrm{sp}(i).\mathrm{nsp}(i)}$ according to (30). Construct the pseudo-variables $\hat{Z}_i$ by plugging $\hat{B}_{\mathrm{sp}(i).\mathrm{nsp}(i)}$ into (33).

(b) Compute the MLE of $B_{i,\text{sp}(i)}$ and $\lambda_i$ in the linear regression (32):

$$\hat{B}_{i,\text{sp}(i)} = Y_i \hat{Z}_i' (\hat{Z}_i \hat{Z}_i')^{-1},$$
$$\hat{\lambda}_i = \frac{1}{n} Y_i (I_N - \hat{Z}_i'(\hat{Z}_i \hat{Z}_i')^{-1} \hat{Z}_i) Y_i'. \quad (34)$$

(c) Use (29) to compute $\hat{B}_{i,\text{nsp}(i)}$ which completes $\hat{B}_i$. Inverting (14) and (15), reconstruct $\hat{\Sigma}^{(r,i)}_{i,-i} = \hat{B}_i \hat{\Sigma}^{(r,i)}_{-i,-i}$, set $\hat{\Sigma}^{(r,i)}_{-i,i}$ equal to the transpose of $\hat{\Sigma}^{(r,i)}_{i,-i}$, and complete $\hat{\Sigma}^{(r,i)}$ by setting $\hat{\sigma}^{(r,i)}_{ii} = \hat{\lambda}_i + \hat{B}_i \hat{\Sigma}^{(r,i)}_{-i,i}$.

(3) Set $\hat{\Sigma}^{(r+1)} = \hat{\Sigma}^{(r,p)}$. Increment the counter $r$ to $r+1$. Go to (2).

The iterations can be stopped according to a criterion such as "the estimate of $\Sigma$ is not changed" (in some pre-determined accuracy).

### 4.5 CONVERGENCE

The key to prove convergence properties of the algorithm in Section 4.4 is to recognize that the algorithm consists of iterated partial maximizations over sections of the parameter space $\mathbf{P}(G)$ (compare Lauritzen 1996, Appendix A.4, and Meng and Rubin 1993). More accurately, we will consider the parameter space

$$\Theta = \{\Sigma \in \mathbf{P}(G) \mid \ell(\Sigma) \geq \ell(\hat{\Sigma}^{(0)})\} \quad (35)$$

which of course contains the global maximizer of $\ell(\Sigma)$. The set $\Theta$ is obviously closed, and under the condition $n \geq p$ necessarily bounded. Thus $\Theta$ is compact.

The section $\Theta_i(\bar{\Sigma}) \subsetneq \Theta$ is defined by

$$\Theta_i = \{\Sigma \in \Theta \mid \Sigma_{-i,-i} = \bar{\Sigma}_{-i,-i}\}. \quad (36)$$

The bijection (31) implies that the algorithm steps (2a)-(2c) maximize the log-likelihood partially over the section $\Theta_i(\hat{\Sigma}^{(r,i-1)})$, i.e.

$$\hat{\Sigma}^{(r,i)} = \arg\max\{\ell(\Sigma) \mid \Sigma \in \Theta_i(\hat{\Sigma}^{(r,i-1)})\}. \quad (37)$$

Hence, the sequence $\ell(\hat{\Sigma}^{(r)})$ is non-decreasing and bounded and thus converges, i.e.

$$\lim_{r \to \infty} \ell(\hat{\Sigma}^{(r)}) = \ell^{(\infty)}. \quad (38)$$

Since $\Theta$ is compact the sequence $\hat{\Sigma}^{(r)}$ must have a convergent subsequence $\hat{\Sigma}^{(r_t)}$ with limit $\hat{\Sigma}^{(\infty)}$. By (37), $\hat{\Sigma}^{(\infty)}$ maximizes the log-likelihood in particular over the section of $\Theta$ defined by fixing all but one single entry $\sigma_{ij}$ of $\Sigma$. This implies that $\hat{\Sigma}^{(\infty)}$ solves the likelihood equations. Moreover, (37) shows that $\hat{\Sigma}^{(\infty)}$ is either a saddle point or a local maximum of the log-likelihood. Finally, (38) implies that $\ell(\hat{\Sigma}^{(\infty)}) = \ell^{(\infty)}$.

If the likelihood equations have only one solution then our algorithm converges to this unique local = global maximum of the likelihood. If there are multiple solutions to the likelihood equations then almost surely no two solutions have the same log-likelihood value. Hence, with probability one, our algorithm constructs a sequence which converges to one of these solutions. The following theorem summarizes our results.

**Theorem 4** *Suppose the sequence $\hat{\Sigma}^{(r)}$ is constructed by the algorithm from Section 4.4. Then all accumulation points of $\hat{\Sigma}^{(r)}$ are saddle points or local maxima of the log-likelihood. Moreover, all accumulation points have the same likelihood value. Thus, with probability one, the sequence $\hat{\Sigma}^{(r)}$ converges to a saddle point or a local maximum of the log-likelihood.*

In practice, the finite accuracy used in computer calculations seems to prevent convergence to a saddle point.

### 4.6 REMARK ON COMPLEXITY

The new algorithm can be restated only in terms of the empirical covariance matrix $S$ defined in (4). For example in (23), $Y_1 \hat{Z}_1' = S_{13} - \hat{\beta}_{32.4} S_{12} - \hat{\beta}_{34.2} S_{14}$ and the remaining quantities can be expressed similarly. Thus, the sample size does not affect the complexity of the algorithm. The complexity of one of the algorithm's pseudo-variable regression steps is dominated by the solution of the systems of $\text{nsp}(i)$ and $\text{sp}(i)$ linear equations in (29) and (34), respectively.

## 5 EXAMPLE DATA

Table 1 presents data on $p = 4$ variables measured on $n = 39$ patients (see Cox and Wermuth 1993, Table 7, and Kauermann 1996, Table 1). If we index the variables in this data set by $W = 1$, $V = 2$, $X = 3$, and $Y = 4$ then the covariance graph model fitted by Kauermann is the one illustrated in Figure 1(a). We use the observed marginal correlations and standard deviations to reconstruct the empirical covariance matrix. Then we fit the model from Figure 1(a) by our new algorithm for ML estimation. Table 2 shows that the ML estimates and Kauermann's dual estimates are very similar in this example.

Table 1: Observed Marginal Correlations and Standard Deviations.

|    | W      | V      | X      | Y    |
|----|--------|--------|--------|------|
| V  | 0.060  |        |        |      |
| X  | −0.460 | 0.042  |        |      |
| Y  | −0.071 | −0.404 | −0.334 |      |
| SD | 5.72   | 92.00  | 7.86   | 2.07 |

Table 2: Marginal Correlations and Standard Deviations from ML (Lower Half & 6th Row) and Kauermann's Dual Estimation (Upper Half & 5th Row).

| ML\dual | W | V | X | Y |
|---|---|---|---|---|
| W |  | 0 | −0.479 | 0 |
| V | 0 |  | 0 | −0.373 |
| X | −0.475 | 0 |  | −0.351 |
| Y | 0 | −0.378 | −0.342 |  |
| $SD_{dual}$ | 5.70 | 91.6 | 7.92 | 2.04 |
| $SD_{ML}$ | 5.72 | 92.0 | 7.93 | 2.05 |

## 6 CONCLUSION/EXTENSIONS

The new algorithm finds ML estimates in covariance graph models using only standard least squares tools. Thus its implementation is straightforward. Moreover, with probability one, the new algorithm converges to a saddle point or local maximum of the likelihood.

Besides the modification discussed in Section 4.2.4, another modification which potentially speeds up our algorithm consists in using multivariate regressions instead of the univariate regressions $(Y_i \mid Y_{-i})$. The multivariate regressions would be of the form $(Y_C \mid Y_{V \setminus C})$ for some subset $C \subseteq V$. If the subset $C$ is complete with respect to $G$, i.e. every pair of vertices in $C$ is adjacent, then the conditional distribution $(Y_C \mid Y_{V \setminus C})$ has the form of a seemingly unrelated regression. ML estimation in seemingly unrelated regression itself requires iterative algorithms but if the current estimate of $\Sigma$ is used as a starting value then a single step in such an algorithm would be sufficient for the extension of our algorithm. Following this idea, one could perform edge-wise updates, i.e. $|C| = 2$, but it might possibly be better to perform updates for cliques $C$.

Finally, since the new algorithm exploits regression techniques, and not directly the likelihood equations, it may lend itself to generalization; for example, to the case of a graphical model for marginal independence in which the variables are discrete.

**Acknowledgements**

NSF grants DMS 9972008 and DMS 0071818.